\documentclass[english]{article}
\usepackage[utf8]{inputenc}
\usepackage{graphicx}
\usepackage[margin=1in]{geometry}
\usepackage{amsmath}
\usepackage{multirow}
\usepackage{booktabs}
\usepackage{subfig}
\usepackage{algpseudocode}
\usepackage{algorithm}
\usepackage{hyperref}
\newcommand{\pluseq}{\mathrel{+}=}

\usepackage{authblk}

\title{Using coarse GPS data to quantify city-scale transportation system resilience to extreme events}
\author[1]{Brian Donovan\thanks{bpdonov2@illinois.edu }}
\author[1]{Daniel B. Work\thanks{dbwork@illinois.edu}}
\affil[1]{Department of Civil and Environmental Engineering, \authorcr University of Illinois at Urbana Champaign}
\begin{document}
  \maketitle
\section*{Abstract }
This article proposes a method to quantitatively measure the resilience of transportation systems using GPS data from taxis.  The granularity of the GPS data necessary for this analysis is relatively coarse; it only requires coordinates for the beginning and end of trips, the metered distance, and the total travel time.  The method works by computing the historical distribution of pace (normalized travel times) between various regions of a city and measuring the pace deviations during an unusual event.  This method is applied to a dataset of nearly 700 million taxi trips in New York City, which is used to analyze the transportation infrastructure resilience to Hurricane Sandy.  The analysis indicates that Hurricane Sandy impacted traffic conditions for more than five days, and caused a peak delay of two minutes per mile.  Practically, it identifies that the evacuation caused only minor disruptions, but significant delays were encountered during the post-disaster reentry process.  Since the implementation of this method is very efficient, it could potentially be used as an online monitoring tool, representing a first step toward quantifying city scale resilience with coarse GPS data.

\section{Introduction}

\subsection{Motivation}
\label{sec:motivation}
In recent years, resilience of city infrastructure has gained a great deal of attention \cite{nchrp777}.  When disasters and other extreme events occur, infrastructure may fail, incurring large human, economic, and environmental costs.  This is especially relevant for transportation infrastructure, since it is crucial for city evacuations and emergency services in post--disaster environments.  Methods are needed to quantitatively monitor the transportation infrastructure in terms of its ability to withstand and recover from such events.  Measuring the performance of city-scale infrastructure with traditional traffic sensors is cost--prohibitive due to relatively high installation costs, but many cities already have taxi fleets equipped with GPS sensors.  Though this analysis could be performed with any GPS data, taxi data is publicly available in some cases.  The New York City dataset used in this analysis gives interesting insights about the performance of infrastructure during Hurricane Sandy and other major events.

The goal of this article is to develop and implement a method for measuring resilience of city-scale transportation networks using only taxi datasets.  The technique is designed with the following characteristics:

\begin{enumerate}
\item \textbf{The method can be applied at the city-scale, or larger.}  Extreme events such as hurricanes have the ability to affect an entire city.  For this reason, it is important to examine impacts at a high-level city view, rather than the level of individual vehicles or streets.
\item \textbf{The method measures network performance quantitatively, in terms of recovery time and peak pace deviations.} Recovery time and peak performance degradation are fairly standard quantities of interest in the resilience literature \cite{aven2011some, haimes2009definition}.  While travel times are a natural performance measure for transportation networks, we instead use \textit{pace}, or travel time per mile.   This normalization accommodates the varied length of taxi trips within a city.
\item \textbf{The method accommodates the inherent variability in traffic conditions and data.}  The available data is full of noise and depends on many unmodeled human factors.  As a result, the method evaluates events that cause statistically significant disruptions, in order to separate the signal from the noise.
\item \textbf{The method is computationally tractable.}  Since taxi trips occur very frequently in large cities, the amount of data available for analysis is large.  In order to be tractible, the computation should be $O(N)$, where $N$ is the number of taxi trips, and ideally require only one pass through the raw data.  Of practical significance, these single-pass algorithms could also be used to process the data in a realtime stream.
\end{enumerate}

\subsection{Related Work}
In recent years, the study of resilience has gained popularity in the systems engineering community.  Haimes \cite{haimes2009complex, haimes2009definition, haimes2011responses} gives a framework for assessing resilience, which focuses on modeling a system and the possible outcomes of various events.  He asserts that a resilient system should suffer only slight degredation during an event, then rapidly recover.  Reed et al. \cite{reed2009methodology} note that the quality of service abruptly drops during an event, then exponentially decays back to typical values.  They suggest that an appropriate resilience measure is the integral of this exponential curve.  Authors in the related field of risk analysis emphasize the importance of unknown factors while assessing resilience \cite{aven2011some, kaplan1981quantitative}.

Though there is no precise consensus on the definition of resilience, \textit{peak disruption} and \textit{recovery time} are consistently discussed quantities.   In other words, peak disruption measures how far the quantity of interest deviates from typical values, and recovery time measures how long it takes to return to typical values. Most of these works also emphasize that resilience must be measured with respect to a given event and quantity of interest.  For example, one case study used the number of functioning nodes in a power grid as the quantity of interest, assessing resilience against hurricanes and minor events \cite{ouyang2012three}.  This article will follow this standard in the sense that it will use GPS data to measure the resilience of a transportation network with respect to specific events.  No claims are made about the overall resilience of the network.

Several authors have proposed quantities of interest for transportation systems.  Omer et al. \cite{omer2013assessing} proposed a method which measures the resilience of a road-based transportation network in terms of travel times between cities.  Chang et al. \cite{chang2001measuring} evaluated a post-earthquake transportation network in terms of accessibility and coverage.  This is partly based on an accessiblilty metric devised by Allen et al. \cite{allen1993accesibility}, which considers travel times between various regions of a city.  Thus, travel time is a standard quantity on which to measure resilience.  This article will use the related quantity of pace, or travel time per mile.

A distinct set of studies use large amounts of data to extract useful information about urban systems.  The work most closely related to resilience is a study by He and Liu \cite{he2012modeling}, which uses loop detector data to measure the effect of the I-35W bridge collapse in Minneapolis in 2007.  Geroliminis et al.  \cite{geroliminis2008existence} use loop detector data, combined with 500 GPS vehicles to extract macroscopic traffic properties from an urban-scale transportation network.  Other works use GPS traces of mobile devices to analyze movement patterns of crowds during typical days and atypical events \cite{calabrese2010geography, calabrese2011real}.  Castro et al. \cite{castro2012urban} present a method for inferring current and future traffic states from taxi GPS data.  Zheng et al. \cite{zheng2011urban} propose a method that tracks taxi trips between various regions of a city and identifies flawed urban planning.  Another study measures temporal patterns in the density of taxi pickups and dropoffs to identify the social function of various city regions \cite{qi2011measuring}.  They point out that unusual output can be used to detect events like holidays.  Chen \cite{chen2012real} specifically focuses on identifying anomalous taxi trajectories, in order to detect fraud or special events. Ferreira et al. \cite{ferreira2013visual} created a graphical querying tool which can be used to count taxi trips between arbitrary geometrical regions as a function of time.  They noted the drop in the frequency of taxi trips during Hurricane Sandy and Hurricane Irene, pointing out that the Irene-related drop was more significant, but the Sandy-related drop was longer lasting.  By examining pace, we confirm that Hurricane Sandy had a longer recovery time, but find the contrasting result that Hurricane Sandy also has a more significant peak disruption.

\subsection{Outline and Contributions}
The contributions of this work are as follows.   In Section~\ref{sec:methodology}, a method is proposed to use taxis as pervasive city-scale resilience sensors.  This method detects unusual events and measures them in terms of peak disruption and recovery time.  It introduces paces between regions of the city as the key performance measure, and it uses the historical pace distribution to detect and measure extreme events. In Section~\ref{sec:Application}, the method is applied to a four-year dataset from New York City to identify and compare properties of events such as Hurricane Sandy.  Of practical significance, the analysis identifies the relative efficiency of the pre-Sandy evacuation, contrasted with the gridlock of post-Sandy reentry. Conclusions and future work is summarized in Section~\ref{sec:conclusion}.  As a technical contribution, all code \cite{github} and data \cite{taxidata} used in this analysis are made publicly available.

\section{Methodology}
\label{sec:methodology}
\subsection{Overview}
The proposed technique to measure city-scale resilience of the transportation network in response to various events by examining taxi trip data is done in three steps.  In section \ref{subsec:feature_extraction}, individual taxi trips are aggregated by origin-destination pairs in order to measure typical paces between various regions of the city.  This aggregation technique makes it possible to extract city-scale features at various points in time, since it is difficult to measure resilience from individual trips.  Section \ref{subsec:typical_behavior} imposes a one-week periodic pattern on the paces, defining the mean and variance of paces for each hour of the week.  Finally, Section \ref{subsec:detect_dev} uses these distributions to quantify how typical or atypical the pace is at a particular point in time.  Atypical paces (e.g., the 5\% most unlikely points) are flagged as events, and they are examined in more detail.

\subsection{Extraction of Time-Series Features from Aggregated Trips}
\label{subsec:feature_extraction}
In the first stage of analysis, trips are grouped by their geographic locations and times of occurrence.  More specifically, the city is divided into a small number, $k$, of large regions.  This allows each taxi trip to be labeled as one of $k^2$ unique origin-destination pairs.  Time is discretized into hours, so a large sample of trips can be gathered at any point in time.  The start zone, end zone, and departure time are used to partition all of trips into subsets. The variable $T_{i,j,t}$ denotes the set of all trips from zone $i$ to zone $j$ at time $t$:

\begin{equation}
T_{i,j,t} = \left\{ r | o(r) \in z(i) , d(r) \in z(j) , \lfloor s(r) \rfloor = t \right\},
\end{equation}
where $o(r)$ is the origin of trip $r$, $d(r)$ is the destination of trip $r$, $z(i)$ is the geographic region of zone $i$, and $\lfloor s(r) \rfloor$ is the start time of trip $r$, rounded down to the hour.  It is assumed that $i$ and $j$ are both in $\{0, 1, \cdots, k-1\}$.  Once these subsets of trips are defined, macroscopic traffic features can be extracted from them.  Of particular interest is the expected travel time between two regions.  However, travel times of individual vehicles between two regions are not uniform, due to the varying lengths of trips that connect the same regions.  Much of this variation can be accounted for by normalizing against distance.  In this way, the \textit{average pace} is computed for each trip subset $T_{i,j,t}$.  Trips are weighted by their distance, since longer trips give more information about the state of traffic.  In this way,  the distance-weighted average pace, $P(i,j,t)$, of taxis from zone $i$ to zone $j$ at time $t$ is computed:

\begin{equation}
\label{eq:mean_pace}
P(i,j,t)
= \frac{\sum\limits_{r \in T_{i,j,t}} l(r)p(r)}{\sum\limits_{r \in T_{i,j,t}} l(r)}
= \frac{\sum\limits_{r \in T_{i,j,t}} l(r)\frac{u(r)}{l(r)}}{\sum\limits_{r \in T_{i,j,t}} l(r)}
= \frac{\sum\limits_{r \in T_{i,j,t}} u(r)}{\sum\limits_{r \in T_{i,j,t}} l(r)},
\end{equation}
where $u(r)$ is the travel time of trip $r$, $l(r)$ is the metered length of trip $r$,  and $p(r) = \frac{u(r)}{l(r)}$ is the pace of trip $r$.  For a fixed value of $t$, all $k^2$ distance-weighted average paces collectively form the \textit{mean pace vector}, $\mathbf{a}$.  This vector is a function of time, and contains the $k^2$ pace values at a particular point in time.  Specifically, the $n$th element of $\mathbf{a}(t)$ is given by
\begin{equation}
\mathbf{a}(t)_n = P\left(\left\lfloor{\frac{n}{k}}\right\rfloor, n \bmod k, t\right),
\end{equation}
where $n \in \{0,1,2,\cdots,k^2-1\}$.

It is desirable to use pace as the performance metric instead of the more traditional measure of vehicle counts, since the goal is to measure traffic conditions during extreme events.  If the flow of vehicles between two regions drops significantly, it is difficult to determine whether this is due to increased congestion or decreased demand.  However, an increase in pace indicates congestion, while a decrease in pace indicates decreased demand.  Although the pace of taxis might be a biased estimate of the pace of all vehicles, logic dictates that if taxi drivers are stuck in traffic jams, so are the other vehicles around them.

\subsection{Identification of City-Scale Typical Behavior}
\label{subsec:typical_behavior}
The mean pace vector, $\mathbf{a}(t)$, has a strongly periodic weekly pattern.  During rush hour, the pace is high, especially in dense downtown regions, and at night the pace is low.  On weekends, the rush hour is less extreme.  However, the mean pace vector has some variance around this periodic pattern, so it is viewed as a distribution conditioned on time.  For example, the mean pace vector for all Tuesdays at 3pm will be slightly different, and significantly different during an unusual event.  To facilitate this grouping, the \textit{reference set} $Q_t$ is defined for all times $t$.  This set contains all of the mean pace vectors which occur at the same point in the periodic pattern as $\mathbf{a}(t)$, except for $\mathbf{a}(t)$ itself.  Intuitively, when deciding how typical the traffic data is at time $t$, that data should not be used as part of the definition of typical.  Since there are 168 hours in a week, the reference set can be defined as

\begin{equation}
\label{eq:reference_set}
Q_t = \{\mathbf{a}(h) | h \equiv t \bmod{168}, h \neq t\}.
\end{equation}

The reference set $Q_t$ makes it possible to compute the expected value of the mean pace vector $\mathbf{\mu}(t)$ as well as the covariance matrix $\mathbf{\Sigma}(t)$.  This covariance matrix is important because it quantifies the noisy day-to-day fluctuations in the mean pace vector, outside of the event at hand, and how the dimensions correlate.  The time-dependent sample mean and covariance matrices can be defined as:

\begin{equation}
\label{eq:mu_sigma}
\begin{array}{l}
\mathbf{\mu}(t) = \frac{1}{|Q_t|}\sum\limits_{\mathbf{a} \in Q_t}\mathbf{a} \\

\mathbf{\Sigma}(t) = \frac{|Q_t|}{|Q_t| - 1}\left(\sum\limits_{\mathbf{a} \in Q_t}\frac{\mathbf{a}\mathbf{a}^\top}{|Q_t|} - \mathbf{\mu}(t)\mathbf{\mu}(t)^\top\right).

\end{array}
\end{equation}

 If an independence assumption is desired, the diagonal components of these matrices can be extracted.  However, it is likely that many of the $k^2$ dimensions of $\mathbf{a}(t)$ are highly correlated, so the full covariance matrix is used for the remainder of the analysis.  For example, trips that start or end in the same region often have highly correlated paces.  Together, $\mathbf{\mu}(t)$ and $\mathbf{\Sigma}(t)$ make it possible to identify unusual mean pace vectors.

\subsection{Detection of Deviations from Typical Behavior}
\label{subsec:detect_dev}
Intuitively, $\mathbf{\mu}(t)$ captures the expected traffic conditions at a particular point in time.  If the observed traffic conditions are significantly far from this expectation, then those conditions are classified as an extreme event.  The covariance matrix $\mathbf{\Sigma}(t)$ is also considered; if there is typically very little deviation from $\mathbf{\mu}(t)$, then a large deviation is even more extreme.  In one dimensional cases, this is typically addressed by standardizing the data via a z-score.  In higher dimensions, the generalized z-score is called the Mahalanobis distance \cite{mahalanobis1936generalized}.  For this analysis, the Mahalanobis distance for an observed mean pace vector is viewed as a function of the time that the observation occurred:

\begin{equation}
M(t) = \sqrt{(\mathbf{a}(t) - \mathbf{\mu}(t))^\top \mathbf{\Sigma}(t)^{-1} (\mathbf{a}(t) - \mathbf{\mu}(t))}.
\end{equation}

\begin{figure}
\begin{centering}
\includegraphics[width=1\columnwidth]{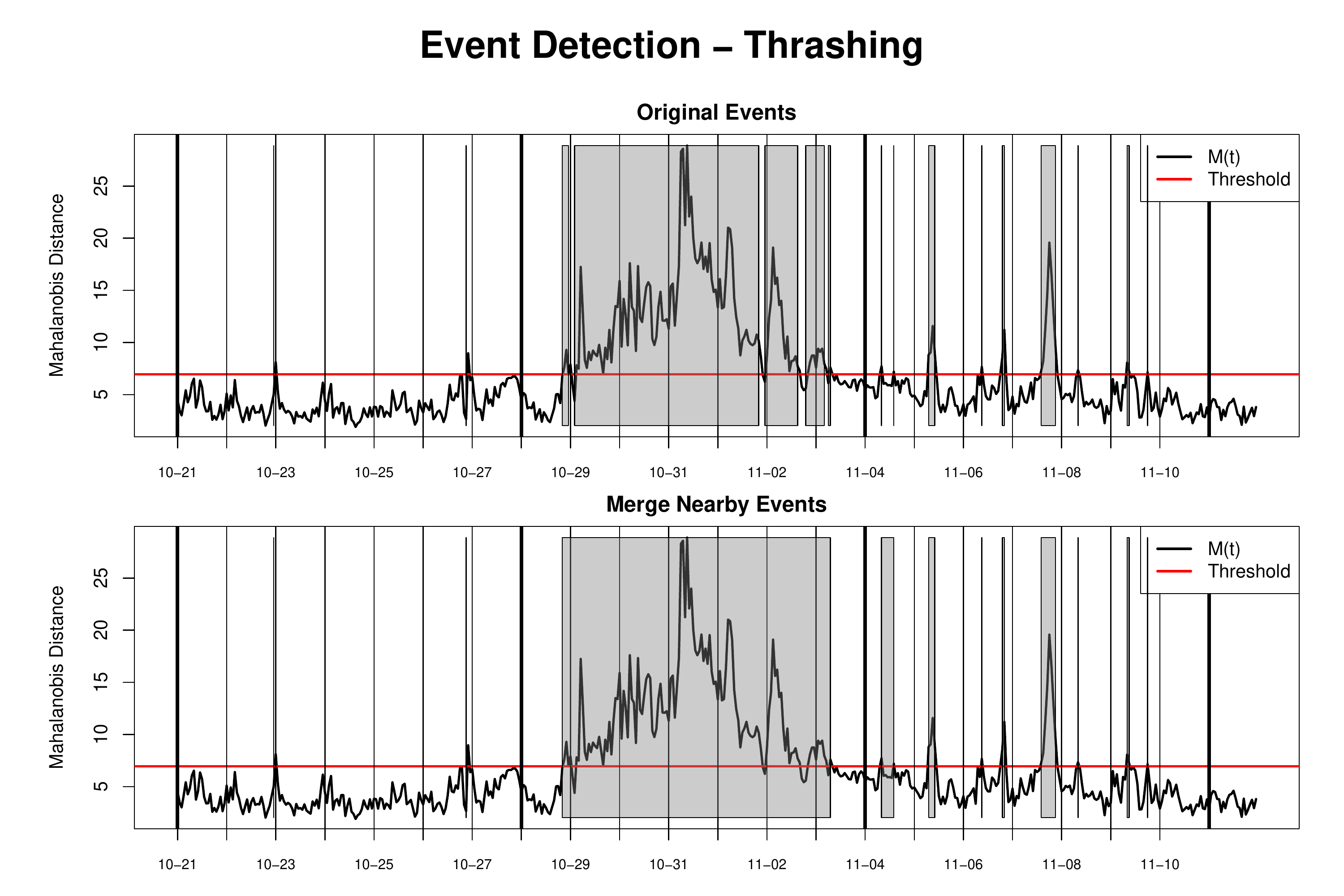}
\par\end{centering}
\caption{Demonstration of event detection.  Events are detected when $M(t)$ goes above the threshold, but thrashing often occurs.  The top graph shows that this thrashing causes events to be divided into several pieces.  For this reason, events with fewer than six hours between them are merged, as shown in the bottom graph. \label{fig:thrashing}}
\end{figure}

This time-dependent Mahalanobis distance serves as an outlier score for observations at various points in time.  Note that it normalizes the deviations in each dimension by the corresponding variances, and also considers correlations between dimensions.  The Mahalanobis distance is a natural way of measuring outliers in multivariate normal data, and it has shown to be useful even when the data is not normal \cite{warren2011use}.  In fact, the multivariate generalization of Chebyshev's inequality gives an upper bound on the probability of observing a Mahalanobis distance greater than some fixed value \cite{navarro2014can}.  In other words, it is unlikely to observe a datapoint with a high Mahalanobis distance, regardless of the distribution.  So, when $M(t)$ rises above a given threshold, an unusual event is detected.  The event is declared complete when $M(t)$ returns to a value lower than the threshold.  In this work, the choice of the threshold is the 95\% quantile of $M(t)$, but this value can easily be lowered to detect smaller events or raised to detect only the most severe events.  The function $M(t)$ is a fairly noisy, which means that it can occasionally \textit{thrash} over the threshold.  In other words, $M(t)$ may rise above the threshold, then immediately drop back below it, effectively breaking the event into two pieces.  To prevent this, consecutive events separated by fewer than six hours are merged.  Figure \ref{fig:thrashing} illustrates this process.

Once the recovery time of an event is computed, other properties can be computed.  For example, it is possible to compute the maximum pace deviation, or the slowest type of trip during the event.  Thus, each event can be described with a set of meaningful statistics.  Comparisons between various events make it possible to describe which types of events the city can easily endure, and where there is room for improvement.  For longer-lasting events like Hurricane Sandy, it is possible to examine different stages of the event in greater detail.

\section{Application to Hurricane Sandy with New York City Taxi Data}
\label{sec:Application}
In this section, the previously described methodology is applied to a dataset of New York City taxi trips.  This dataset, which was obtained through a \textit{Freedom of Information Law} (FOIL) request, covers four years of operation and details nearly 700 million trips.  Many events are detected within this four year span and compared quantitatively.  Special attention is given to Hurricane Sandy and some interesting properties are discovered.

\subsection{The Dataset}
\label{subsec:dataset}
The data used in this analysis takes the form of a large table where each row represents a single taxi trip.  Table \ref{tab:samp_data} gives a small sample of this data. Note that this data format is the minimum amount of information required to perform the analysis.  Other datasets may contain, for example, periodic GPS updates, but this is at least as much information as the New York City data. As there are several entries per second for four years, the raw data takes up about 116GB in text CSV format.  We have made this large dataset publicly available \cite{taxidata}.

\setlength{\tabcolsep}{3.5pt}
\begin{table}
\centering
\bgroup
\def\arraystretch{2.5}
\begin{tabular}{cccccccc}
\toprule
\shortstack{pickup \\ datetime}    & \shortstack{dropoff \\ datetime}   & \shortstack{duration \\ (sec)} & \shortstack{distance \\ (mi)} & \shortstack{pickup \\ lon} & \shortstack{pickup \\ lat} & \shortstack{dropoff \\ lon} & \shortstack{dropoff \\ lat} \\
\midrule
\shortstack{2013-05-01 \\ 00:02:11 } & \shortstack{2013-05-01 \\ 00:14:28 } & 737      & 2.9      & -74.00            & 40.74            & -74.01             & 40.71             \\
\shortstack{2013-05-01 \\ 00:02:12 } & \shortstack{2013-05-01 \\ 00:12:31 } & 618      & 1.8      & -74.00            & 40.73            & -73.98             & 40.72             \\
\shortstack{2013-05-01 \\ 00:02:12 } & \shortstack{2013-05-01 \\ 00:07:39 } & 326      & 1.3      & -73.97            & 40.76            & -73.96             & 40.77             \\

\shortstack{2013-05-01 \\ 00:02:13 } & \shortstack{2013-05-01 \\ 00:04:35 } & 141      & 0.6      & -73.99            & 40.75            & -74.00             & 40.75             \\

\shortstack{2013-05-01 \\ 00:02:14 } & \shortstack{2013-05-01 \\ 00:04:09 } & 115      & 0.5      & -73.98            & 40.75            & -73.99             & 40.74             \\
\bottomrule
\end{tabular}
\caption{A small subset of the data used in this analysis.  Each row corresponds to an occupied taxi trip.}
\label{tab:samp_data}
\egroup
\end{table}

Note that this data only records trips where the taxi is occupied by a passenger.  Non-occupied trips are not recorded. The dataset also contains a large number of errors.  For example, there are several trips where the reported meter distances are significantly shorter than the straight-line distance, violating Euclidean geometry.  Additionally, many trips report GPS coordinates of (0,0), or contain impossible distances, times, or velocities.  All of these types of obvious errors are discarded and account for roughly 7.5\% of all trips.

\begin{figure}
\begin{centering}
\includegraphics[width=0.8\columnwidth]{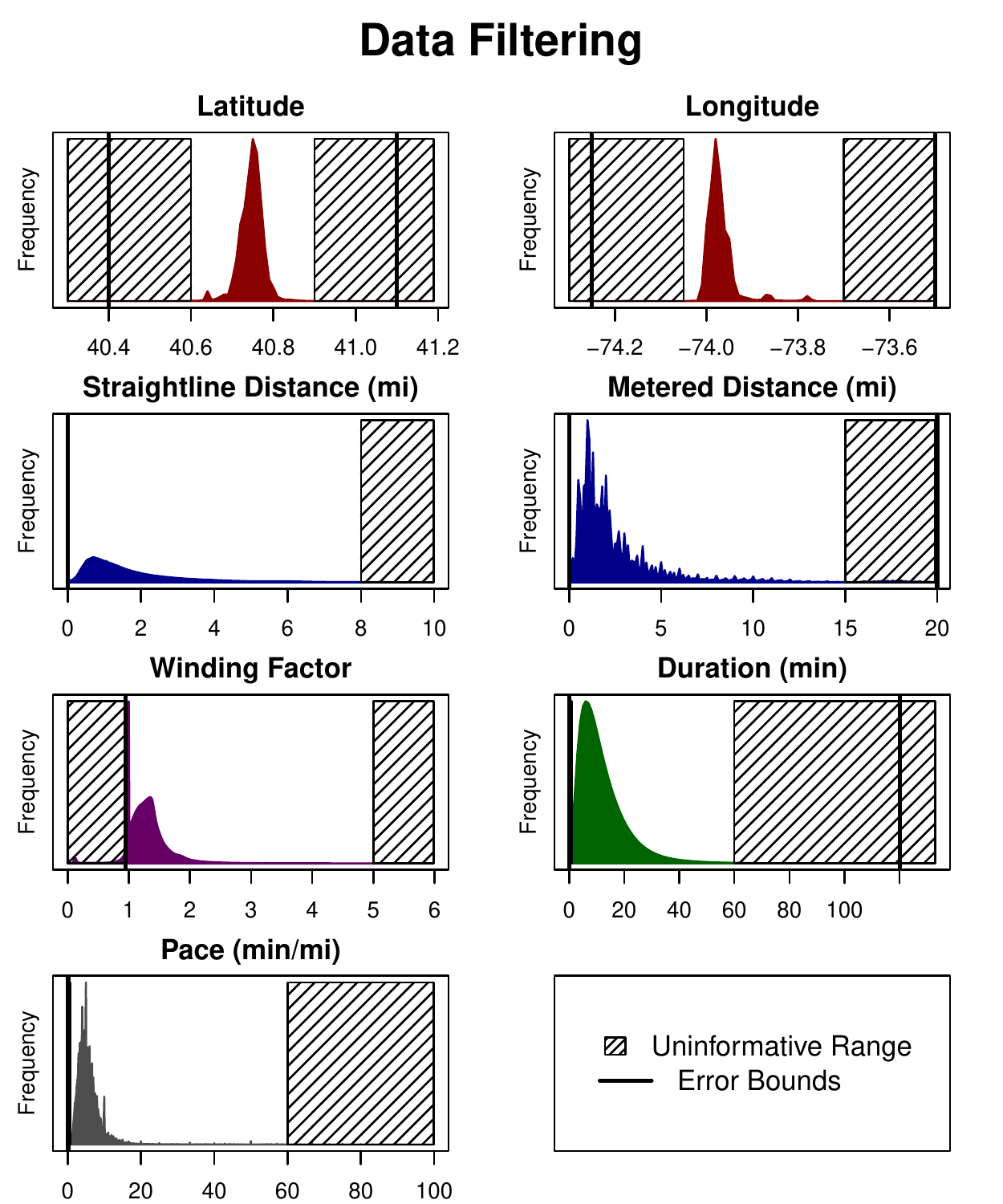}
\par\end{centering}
\caption{Distributions of individual features of taxi trips.  Simple thresholds are used to filter trips that contain errors, or are otherwise uninformative.  Note that the \textit{straightline distance} is the Euclidean distance between start and end coordinates, while the \textit{metered distance} is the value reported by the taximeter.  The \textit{winding factor} is the metered distance divided by the straightline distance.  A winding factor less than 1 is geometrically impossible, and a large value indicates that the taxi did not proceed directly to its destination. \label{fig:error_filter}}
\end{figure}

After removing errors, the dataset is then filtered to remove data outside of the scope of the analysis.  For example, there are many trips which start in Midtown, travel over 50 miles, then end less than a block from their starting points.  These trips are entirely possible, but unlikely to be representative of Midtown-to-Midtown trips because they likely drove many miles in other areas.  This filter is implemented by thresholding the \textit{winding factor}, or metered distance over straight-line distance.  Trips which last less than 60 seconds are also unlikely to give accurate pace estimates because the initial non-driving time becomes more important.  These types of trips are also removed, accounting for roughly 4\% of the original data.  Figure \ref{fig:error_filter} shows histograms of all trip features considered for filtering, as well as the thresholds used for invalid data.  Additionally, the entire months of August and September 2010 were discarded due to a high number of errors.

\subsection{Computational Issues}
Due to the size of the dataset, an efficient software implementation of the analysis is crucial.  This section discusses the algorithmic and practical aspects of the analysis, using the New York City taxi dataset as an example.  In this way, concrete figures can be used for quantities like runtime or data size.  More general concepts like time complexity do not depend on the dataset.

The first step described in Section \ref{subsec:feature_extraction} is the most computationally expensive.  All of the 697,622,444 individual trips are aggregated into 35,064 mean pace vectors - one for each hour in the four-year dataset.  Since the trip data is sorted chronologically, it is possible to compute these mean pace vectors in a single pass.  Recall from \eqref{eq:mean_pace} that the mean pace computation involves the sum of trip durations and the sum of trip distances.  Thus, these two sums are initialized to zero for each of the 16 types of trips.  Each time a trip is read from the file, the relevant sums are incremented.  The error filtering from Section \ref{subsec:dataset} can also be performed at this stage, so an additional pass of the dataset is not required.  When the start hour of the current trip (rounded) is greater than the start hour of the previous trip, the sums are complete for the previous hour. The mean pace vector is computed by division and output, then the sums are reset to zero.  Thus, the computation is one large loop over the entire dataset.  A short pseudocode is given in Algorithm~\ref{Alg:fast}.  Note that NUM\_TYPES is 16, since there are four regions.

\begin{algorithm}
\caption{Online Mean Pace Vector Extraction}\label{Alg:fast}
\begin{algorithmic}
\State prev\_hour $:= -1$         \Comment{Start at beginning of time}
\State sum\_duration $:=$ zeros(NUM\_TYPES) \Comment{Initialize sums to 0}
\State sum\_distance $:=$ zeros(NUM\_TYPES) \Comment{Initialize sums to 0}
\ForAll{trip $\in$ chronological\_trips} \Comment{Loop over all trips}
    \While{trip.hour $>$ prev\_hour}\Comment{If previous hour is complete:}
        \State output$\left(\text{prev\_hour},\frac{\text{sum\_duration}}{\text{sum\_distance}}\right)$ \Comment{Output mean pace vector}
        \State sum\_duration $:=$ zeros(NUM\_TYPES) \Comment{Reset sums to 0}
        \State sum\_distance $:=$ zeros(NUM\_TYPES) \Comment{Reset sums to 0}  
        \State prev\_hour$\pluseq$ 1                 \Comment{Advance to next hour}
    \EndWhile
    \If{trip.isValid()} \Comment{Data filtering}
        \State $i \gets$ category(trip.pickup, trip.dropoff) \Comment{Determine trip type}
        \State sum\_duration[$i$] $\pluseq$ trip.duration \Comment{Update distance sum}
        \State sum\_distance[$i$] $\pluseq$ trip.distance \Comment{Update duration sum}
    \EndIf
\EndFor
\end{algorithmic}
\end{algorithm}

Since each trip is accessed only once, the computation is $O(N)$, where $N$ is the total number of trips.  The computation of each hour timeslice is independent, making it possible to employ parallel processing if the data is partitioned ahead of time.  The analysis was implemented in Python (source code available at \cite{github}) and run on an 8-core 2.5GHz machine with 24GB of RAM.  The extraction of all 35,064 mean pace vectors took about 75 minutes, using roughly 40MB of RAM for each of the eight processes.  The fact that the runtime is much shorter than the real timespan of the dataset combined with the single-pass property means that this preprocessing could be performed in realtime.  In other words, this system could realistically collect trips as they occur, update the relevant sums, then output the mean pace vector at the end of the hour.

The remaining computations involve mean pace vectors instead of raw trip data.  They also have linear time complexity and are much faster than the preprocessing.  Recall from \eqref{eq:reference_set} and \eqref{eq:mu_sigma} that, at a particular hour, the mean and covariance need to be computed for \textit{all hours in the periodic pattern except that hour}.  The naive implementation of this calculation has a quadratic time complexity, since each mean pace vector much be compared against every other mean pace vector in the group.  However, it is possible to compute all of these quantities in linear time.  Instead of directly computing the mean of all values except $\mathbf{A}(t)$, the sum of all values including $\mathbf{A}(t)$ is computed up front.  Then, in the loop, $\mathbf{A}(t)$ is subtracted from this sum.  Formally, the \textit{inclusive reference set}, $Q_{t+}$, is defined in a similar way to \eqref{eq:reference_set}, except that it includes the mean pace vector $\mathbf{A}(t)$.  In other words,

\begin{equation}
Q_{t+} = \{\mathbf{A}(h) | h \equiv t \bmod{168}\} = Q_t \cup \{\mathbf{A}(t)\}.
\end{equation}

Unlike the reference set from \eqref{eq:reference_set}, the inclusive reference set is identical for values of $t$ that occur at the same point in the periodic pattern.  Thus, $Q_{t+}$ and the sum of all vectors in $Q_{t+}$ only need to be computed once.  To compute the sum of all vectors in $Q_{t+}$ \textit{except} $\mathbf{A}(t)$, it is sufficient to subtract $\mathbf{A}(t)$ from this sum.  Thus, the mean computation can be written as

\begin{equation}
\mathbf{\mu}(t) = \frac{1}{|Q_t|}\sum\limits_{\mathbf{A} \in Q_t}\mathbf{A} = \frac{1}{|Q_{t+}| - 1}\left(\left(\sum\limits_{\mathbf{B} \in Q_{t+}}\mathbf{B}\right) - \mathbf{A}(t)\right).
\end{equation}

A similar technique is used for the sum of outer products in the covariance computation.  This method avoids redoing most of the addition in each iteration, allowing for a significant improvement on large datasets.  Once $\mathbf{\mu}(t)$ and $\mathbf{\Sigma}(t)$ are computed, $M(t)$ can be computed in constant time.  Thus, the entire operation runs in linear time.  On the same machine, this computation ran in less than 10 seconds, producing the timeseries of $M(t)$.  Again, this operation would be feasible in a real-time system.  However, it is worth noting that it may be desirable to re-generate old values of $M(t)$ in light of new information.

Once $M(t)$ is generated, the event detection described in Section \ref{subsec:detect_dev} can also be performed in linear time.  Events and spaces between events are stored as a linked list, where each node contains the start time and end time.  Scanning through $M(t)$ chronologically, a new node in the linked list is generated each time $M(t)$ crosses above or below the threshold.  Then, to remove short spaces between events, this linked list is iterated upon.  Each time a non-event node of less than the desired duration is discovered, that node and its two neighbors are replaced with one larger node.  On the same machine as the previous computations, it took less than one second to perform the event detection.

\subsection{Extraction of Pace Features}
\begin{figure}
\begin{centering}
\includegraphics[width=0.8\columnwidth]{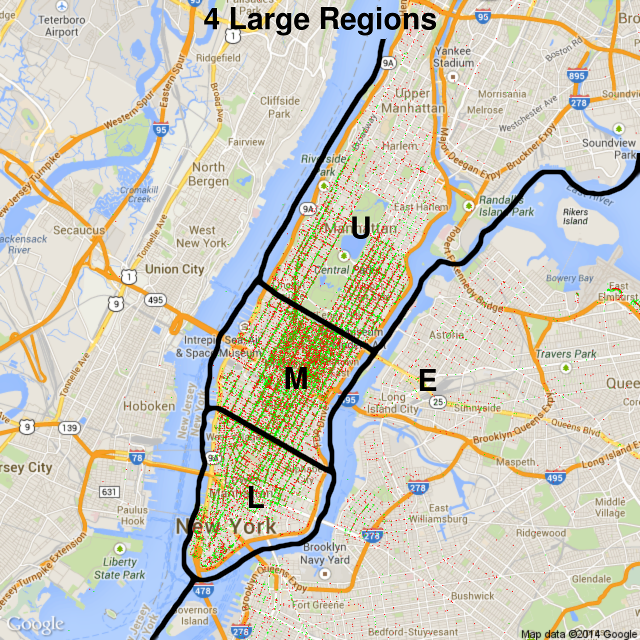}
\par\end{centering}
\caption{Division of New York City into four large regions denoted \textit{U,M,E,} and \textit{L}.  A random sample of 0.01\% of the taxi trips in 2012 are shown.  Pickup locations are marked in green, and the corresponding dropoffs are marked in red.  The majority of trips occur in Manhattan, with  especially high concentration in the Midtown region.\label{fig:regions}}
\end{figure}

The map of New York City is first split into four large regions, shown in Figure \ref{fig:regions}.  For the remainder of the analysis, the zones will be referred to in the following way: \textit{Upper Manhattan} (U), \textit{Midtown} (M), \textit{Lower Manhattan} (L), and \textit{East of the Hudson River} (E).  Note that the Eastern region is connected only by bridges and tunnels and thus problems with this infrastructure will tend to increase travel times between this region and others.  Specifically relevant to Hurricane Sandy is the Lower Manhattan region, since it experienced severe flooding and power outages.  Choosing four large regions in this way satisfies the first goal outlined in Section \ref{sec:motivation} because it defines meaningful city-scale properties.  Instead of looking at every street in New York under a microscope, it defines large areas with key geographic and infrastructural properties.  The travel times between these regions reflect the overall performance of city-scale transportation infrastructure.  It is worth noting that the methodology allows for an arbitrary choice of regions.  This implementation simply chooses zones which are useful for detecting the types of events that occur in New York City.

\begin{figure}
\begin{centering}
\includegraphics[width=1\columnwidth]{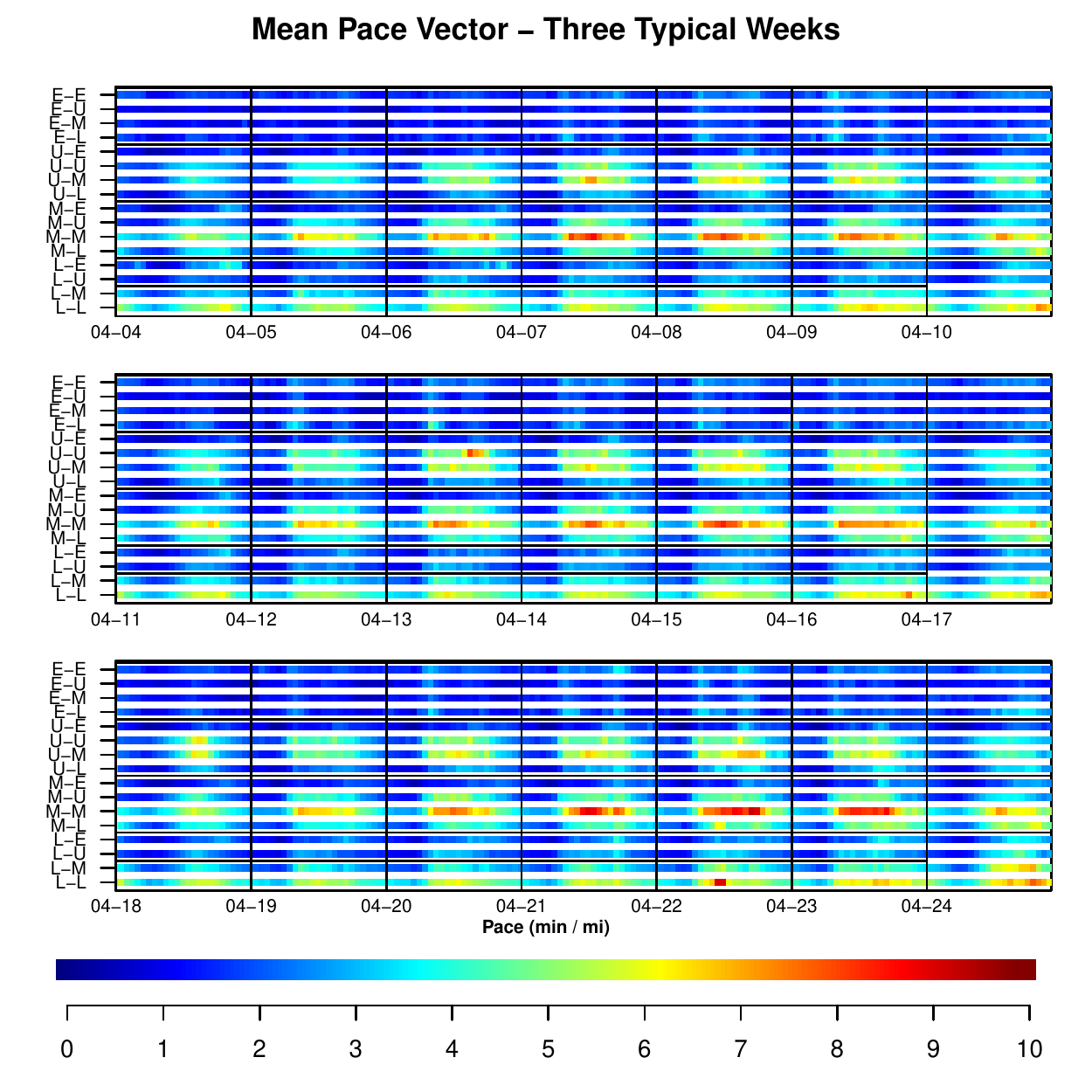}
\par\end{centering}
\caption{The mean pace vector, $\mathbf{a}(t)$ for three typical weeks, starting on April 4, 2010.  A periodic pattern is observable, with high paces during rush hour. \label{fig:mean_pace_vector}}
\end{figure}

Recall that a taxi can take one of 16 possible trips between these regions.  Aggregating these trips by type and hour as in Section \ref{subsec:feature_extraction} produces the 16-dimensional mean pace vector, $\mathbf{A}(t)$, at all points in time.  Figure \ref{fig:mean_pace_vector} shows three typical weeks of mean pace vectors, revealing the expected periodic pattern.

\subsection{Analysis of Events}
As detailed in Section \ref{subsec:typical_behavior}, the expected behavior is generated for all times $t$ according to $\mathbf{\mu}(t)$ and $\mathbf{\Sigma}(t)$.  An interesting way to view the mean pace vector $\mathbf{A}(t)$ is by standardizing it, element by element, producing the \textit{standardized pace vector}.  The $i$th element of this vector is given by

\begin{equation}
\mathbf{S}(t)_i = \frac{\mathbf{A}(t)_i - \mathbf{\mu}(t)_i}{\sqrt{\mathbf{\Sigma}(t)_{i,i}}}.
\end{equation}

\begin{figure}
\begin{centering}
\includegraphics[width=1\columnwidth]{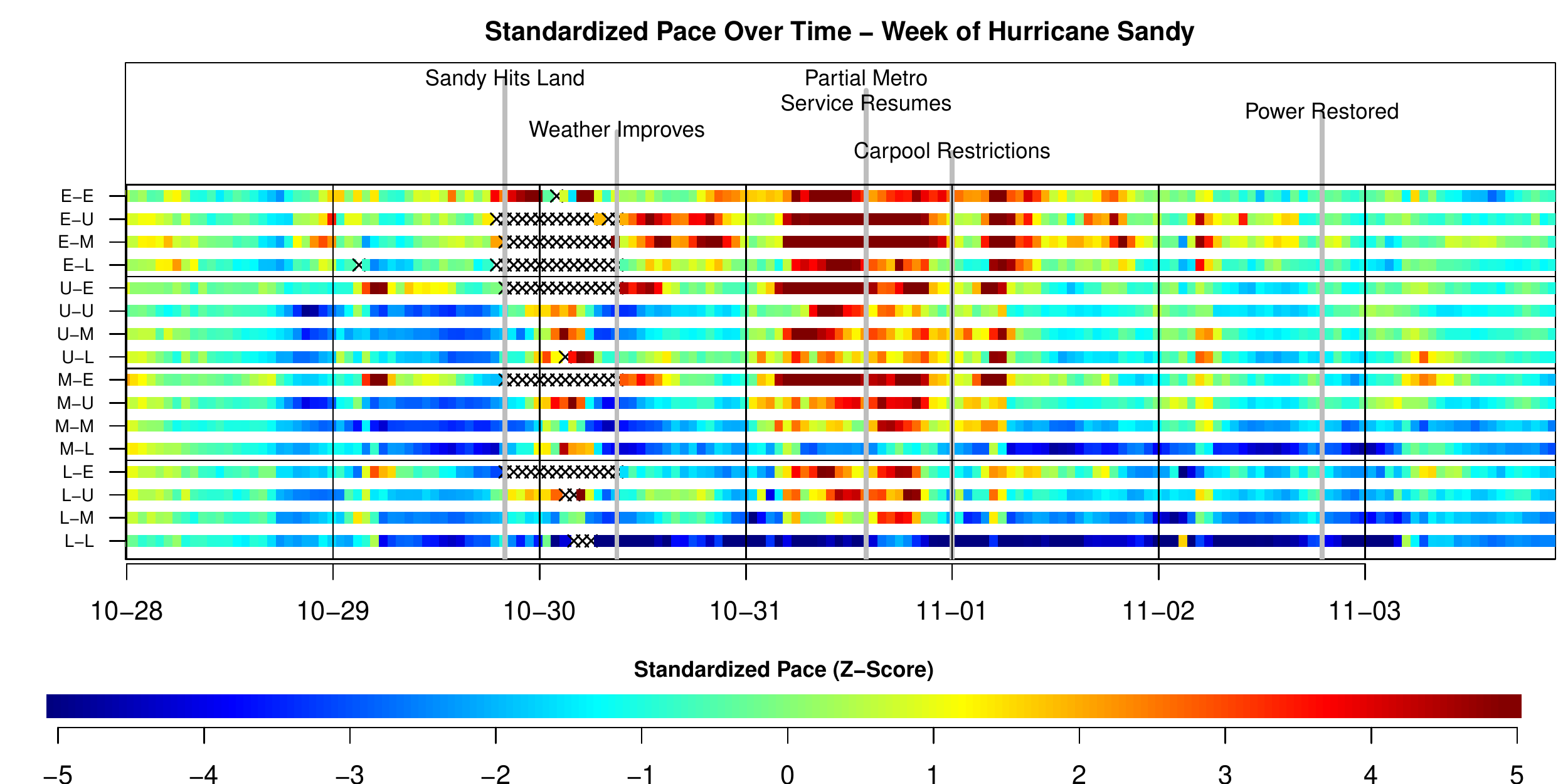}
\par\end{centering}
\caption{The standardized pace vector during the week of Hurricane Sandy, 2012.  Labels are included to show the times of specific phases of the event \cite{rudincenter2012sandy}.  An average week would have values of zero everywhere, but significant deviations are shown during the week of Hurricane Sandy.  Missing data (hours where there are less than five occurrences of a given trip) are marked with black Xs.\label{fig:std_pace_vector}}
\end{figure}

Intuitively, the standardized pace vector tells how many standard deviations away from the mean the pace of each category of trips is at time $t$.  In other words, it is possible to identify the trips that are going slower or faster than expected, and how significant this difference is.  Figure \ref{fig:std_pace_vector} shows the standardized pace vector during the week of Hurricane Sandy.  This figure gives some intuition on the behavior of various regions of the city during and after the hurricane.  It also includes labels indicating the occurrences of various phases of the event, obtained from a post-Hurricane Sandy study from NYU \cite{rudincenter2012sandy}.  Standardizing each origin-destination pace separately allows for additional insight beyond the Mahalanobis distance.

The most notable finding is that the slowest traffic occurred on Wednesday October 31st, almost two days after the hurricane struck land.  On this day, some airports, buses, and commuter rails attempted to resume normal service, but much of the infrastructure was still damaged \cite{rudincenter2012sandy}. It is even more surprising that Midtown-to-Lower Manhattan and Lower Manhattan-to-Lower Manhattan travel times are significantly \textit{lower} than expected during this time.  The pace of these trips remains almost five standard deviations below the mean until Saturday the third, despite the severe flooding and power outages in Lower Manhattan.  The fact that a hurricane can actually make traffic move faster in some areas of the city indicates that the usage of the infrastructure changed.  It is likely that the hurricane decreased demand on the transportation network in Lower Manhattan until the infrastructure began to recover.

\begin{figure}
\begin{centering}
\includegraphics[width=1\columnwidth]{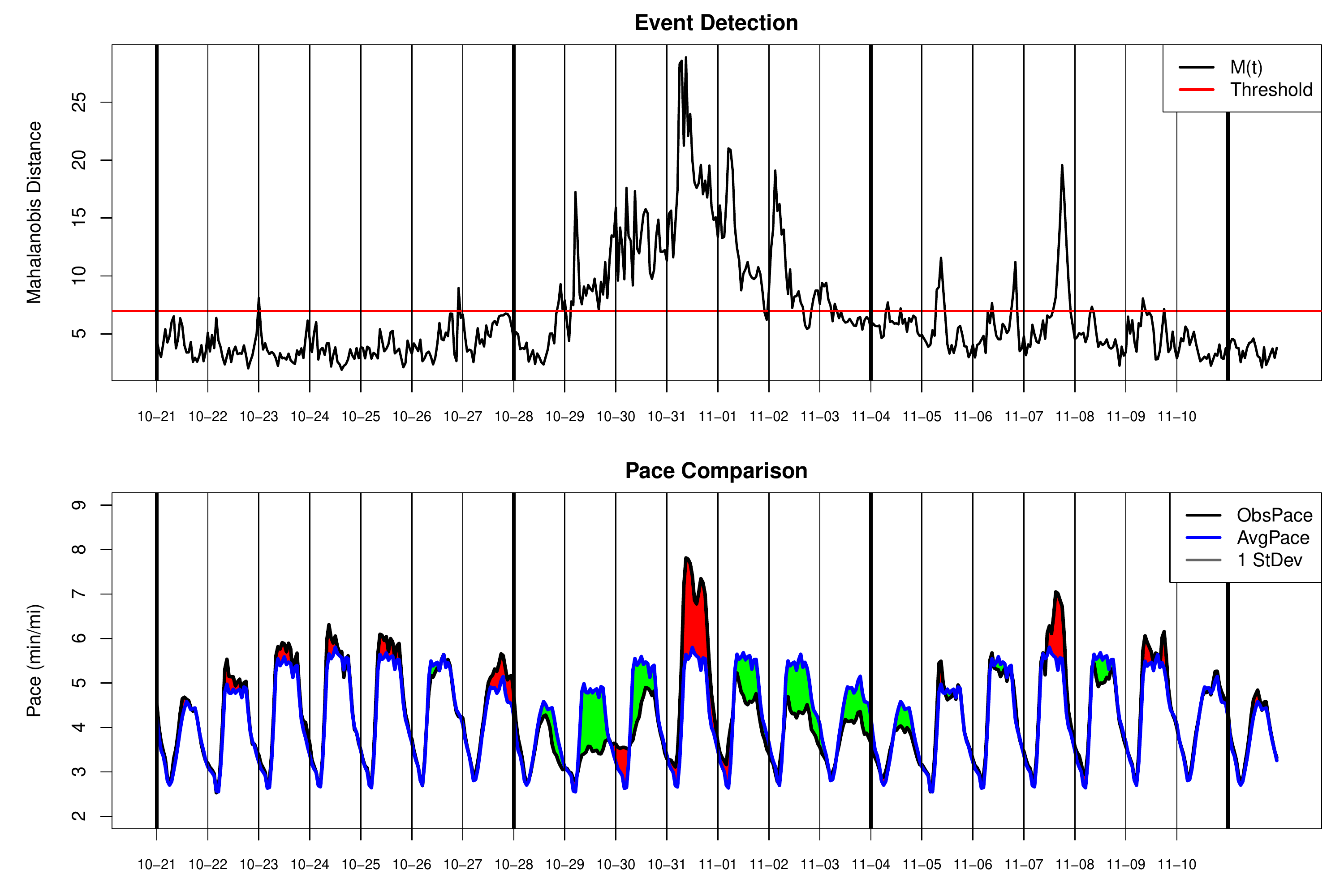}
\par\end{centering}
\caption{Probabilistic detection and measurement of the event Hurricane Sandy. The Mahalanobis distance, $M(t)$,  is plotted in the top figure and events are detected when it goes below the threshold.  For comparison, the average pace of all taxis in the city is plotted below and compared to the expected value.  Green areas indicate that travel times are low, but red indicates that they are unusually high. \label{fig:mahal}}
\end{figure}

This standardized pace vector gives a meaningful interpretation of unusual travel times between various regions of the city, but it fails to account for correlations between these typical travel times i.e., the off-diagonal elements of $\mathbf{\Sigma}(t)$.  In contrast, the Mahalanobis distance $M(t)$ considers the full covariance matrix.  As described in Section \ref{subsec:detect_dev}, events are detected when $M(t)$ goes above a threshold for a significant period of time.  Figure \ref{fig:mahal} shows this process, along with the average pace of all taxis.  Table \ref{table:events} shows the top ten events, sorted by duration.  At the top of the list is Hurricane Sandy, taking over five and a half days for travel times to return to normal.  This is over three times the recovery time of Hurricane Irene.  This agrees with the results of \cite{ferreira2013visual}, which showed that the total number of Manhattan taxi trips returned to normal more quickly during Hurricane Irene than Hurricane Sandy.  At its worst, Sandy added over two minutes to each mile driven by taxis in the city, while Irene added less than forty seconds.  This is in contrast to the results of \cite{ferreira2013visual}, which showed that the peak drop in the number of taxi trips was greater during Hurricane Irene.  The blizzard of December 2010, while shorter, added four minutes of travel time to each mile at its peak.

It is difficult to evaluate the accuracy of the results in Table \ref{table:events}, since the true severity of each event is not known.  If a training set of events is available, one could raise or lower the detection threshold until the desired balance between type I and type II errors is reached.

\begin{table}
\centering
\begin{tabular}{ c  c  c  c  c  c }
\toprule

Event           & Start Time          & \shortstack{Duration \\ (hours)}     & \shortstack{Max \\ (min/mi)}            & \shortstack{Min \\ (min/mi)}            & Worst Trip     \\
\midrule
Sandy     &     2012-10-28 21:00:00     &     132     &     2.25     &     -1.6     &     E $\rightarrow$ M \\
Blizzard     &     2010-12-26 13:00:00     &     112     &     4.41     &     0.33     &     M $\rightarrow$ M \\
Blizzard     &     2011-01-31 08:00:00     &     49     &     2.04     &     0.34     &     E $\rightarrow$ E \\
Irene     &     2011-08-27 13:00:00     &     43     &     0.64     &     -1.66     &     E $\rightarrow$ E \\
Unknown     &     2013-10-12 03:00:00     &     33     &     1.09     &     0.08     &     E $\rightarrow$ L \\
Blizzard     &     2013-02-08 06:00:00     &     26     &     1.54     &     -0.58     &     E $\rightarrow$ E \\
Blizzard     &     2010-02-10 06:00:00     &     24     &     0.67     &     -1.01     &     E $\rightarrow$ E \\
New Years     &     2012-12-31 15:00:00     &     20     &     1.42     &     -2.66     &     E $\rightarrow$ M \\
Unknown     &     2011-09-09 08:00:00     &     19     &     1.66     &     0.35     &     U $\rightarrow$ U \\
Blizzard     &     2011-01-28 02:00:00     &     18     &     2.57     &     0.49     &     L $\rightarrow$ L \\

\bottomrule
\end{tabular}
\caption{Comparison of New York City transportation infrastructure resilience to the 10 longest events.  The duration in hours, and the maximum/minimum pace deviation in minutes/mile is given for each event.  Note that a positive number indicates a delay while a negative indicates a decreased pace.  The final column indicates which of the 16 trips most frequently had the highest standardized pace during the event.  Labels for events (the first column) are determined manually (cf. \cite{searchengine}).}
\label{table:events}
\end{table}

\section{Conclusion}
\label{sec:conclusion}
This analysis has shown that it is possible to detect and measure the effects of unusual events on transportation infrastructure using only taxi GPS data.  This is a first step toward assessing and improving city-scale resilience.  Of key importance, the method is extremely low cost, because it does not require the installation of any additional sensors.  This method proposes computing \textit{origin-destination paces}, or average travel time per mile between various regions of the city.  The effects of various events are quantified by the sizes and durations of pace deviations from typical values.  Importantly, this measurement considers the typical statistics of traffic conditions, so significant events can be distinguished from random day-to-day fluctuations.

The proposed method is applied to a dataset from New York City, and Hurricane Sandy is analyzed in detail.  The analysis shows this was the longest event in the four year dataset, and one of the most severe in terms of peak pace deviation.  At its worst, Hurricane Sandy caused over two minutes of delay per mile, but actually resulted in \textit{faster} traffic for most of its duration.  Most interestingly, the spike in delay occurred two days after the hurricane struck, as many residents migrated back into the city.  This re-entry process was extremely slow when compared to the evacuation process before the hurricane, suggesting that more traffic management might be necessary following an event.  The analysis of an extreme event like Hurricane Sandy demonstrates the ability of the proposed method to capture and describe atypical city-scale properties of the transportation network.

\section{Future Work}
This research is ongoing, and leaves several opportunities for improvement.  For example, in section \ref{subsec:feature_extraction}, regions are chosen manually.  Naturally, one may ask how the results will change when different regions are chosen, or when a different number of regions are used.  This question can be answered empirically by trying several different partitioning schemes.  It may be possible to automatically define regions via clustering as in \cite{ji2012spatial}.  On a related note, more GPS data is clearly required when more regions are used, in order to gain an adequate sample for each origin-destination pair.

It is also possible to apply the outlier-detection methods to other types of paces.  For example, instead of measuring paces between various origin-destination zones, one may desire to compute approximate paces on each link of the network graph.  Algorithms exist which can estimate link travel times, for example \cite{hunter2009path} and \cite{santi2014quantifying}, but they are computationally expensive.  If the same outlier-detection methods are applied to link-level pace data, it is possible to examine whether such a heavy computation is necessary.  If the results are unchanged, the simpler method presented in this article may be sufficient.

\section*{Acknowledgments}

This work was supported by the National Science Foundation under Grant No.  CNS-1308842.

\bibliographystyle{ieeetr}
\bibliography{DonovanWork2015}

\end{document}